\documentclass[sigconf, screen]{acmart}
\renewcommand\footnotetextcopyrightpermission[1]{} 
\usepackage{algorithm}
\usepackage{algorithmic}
\usepackage[table]{xcolor}
\usepackage{makecell}

\usepackage[dvipsnames]{xcolor}
\NewDocumentCommand{\zihao}{mO{}}{%
  \textcolor{Orange}{\textsuperscript{\textit{zihao}}\textsf{\textbf{\small[#1]}}}%
}

\NewDocumentCommand{\lifu}{mO{}}{%
  \textcolor{red}{\textsuperscript{\textit{Lifu}}\textsf{\textbf{\small[#1]}}}%
}

\newcommand{\methodname}{\textsc{GLANCE}}

\AtBeginDocument{%
  }

\setcopyright{acmlicensed}
\copyrightyear{2026}
\acmYear{2026}
\acmDOI{XXXXXXX.XXXXXXX}
\acmConference[MM '26]{Make sure to enter the correct conference title from your rights confirmation email}{Preprint, Under review}

\begin{document}

\title{GLANCE: A Global–Local Coordination Multi-Agent Framework for Music-Grounded Non-Linear Video Editing}

\author{Zihao Lin}
\affiliation{%
  \institution{UC Davis}
  \city{Davis}
  \state{California}
  \country{USA}
}

\author{Haibo Wang}
\affiliation{%
  \institution{UC Davis}
  \city{Davis}
  \state{California}
  \country{USA}
}

\author{Zhiyang Xu}
\affiliation{%
  \institution{UC Davis}
  \city{Davis}
  \state{California}
  \country{USA}
}

\author{Siyao Dai}
\affiliation{%
  \institution{Fudan University}
  \city{Shanghai}
  \country{China}
}

\author{Huanjie Dong}
\affiliation{%
  \institution{UC Davis}
  \city{Davis}
  \state{California}
  \country{USA}
}

\author{Xiaohan Wang}
\affiliation{%
  \institution{Stanford University}
  \city{Palo Alto}
  \state{California}
  \country{USA}
}

\author{Yolo Y. Tang}
\affiliation{%
  \institution{University of Rochester}
  \city{Rochester}
  \state{New York}
  \country{USA}
}

\author{Yixin Wang}
\affiliation{%
  \institution{Stanford University}
  \city{Palo Alto}
  \state{California}
  \country{USA}
}

\author{Qifan Wang}
\affiliation{%
  \institution{Meta AI}
  \city{Menlo Park}
  \state{California}
  \country{USA}
}

\author{Lifu Huang}
\affiliation{%
  \institution{UC Davis}
  \city{Davis}
  \state{California}
  \country{USA}
}

\renewcommand{\shortauthors}{Zihao et al.}

\begin{abstract}
Music-grounded mashup video creation is a challenging form of video non-linear editing, where a system must compose a coherent timeline from large collections of source videos while aligning with music rhythm, user intent, story completeness, and long-range structural constraints. Existing approaches typically rely on fixed pipelines or simplified retrieval-and-concatenation paradigms, limiting their ability to adapt to diverse prompts and heterogeneous source materials. In this paper, we present \textbf{\methodname}, a \emph{global-local coordination multi-agent framework} for music-grounded nonlinear video editing. \methodname{} adopts a bi-loop architecture for better editing practice: an outer loop performs long-horizon planning and task-graph construction, and an inner loop adopts the "Observe-Think-Act-Verify" flow for segment-wise editing tasks and their refinements. To address the cross-segment and global conflict emerging after subtimelines composition, we introduce a dedicated global-local coordination mechanism with both preventive and corrective components, which includes a novelly designed context controller, conflict region decomposition module and a bottom-up dynamic negotiation mechanism. To support rigorous evaluation, we construct \textbf{MVEBench}, a new benchmark that factorizes editing difficulty along task type, prompt specificity,
and music length, and propose an \emph{agent-as-a-judge} evaluation framework for scalable multi-dimensional assessment. Experimental results show that \methodname{} consistently outperforms prior research baselines and open-source product baselines under the same backbone models. With GPT-4o-mini as the backbone, \methodname{} improves over the strongest baseline by 33.2\% and 15.6\% on two task settings, respectively, with particularly strong gains on more challenging long-horizon subsets. Human evaluation further confirms the quality of the generated videos and validates the effectiveness of the proposed evaluation framework. Code will be available at \url{https://github.com/ZihaoLinQZ/GLANCE-Video-Editing-Agent}.
\end{abstract}

\ccsdesc[500]{Computing methodologies~Multi-agent planning}

\keywords{Non-Linear Video Editing, Video Understanding, Multimodal Large Language Models, Multimodal Agentic AI}
\begin{teaserfigure}
  \includegraphics[width=\textwidth]{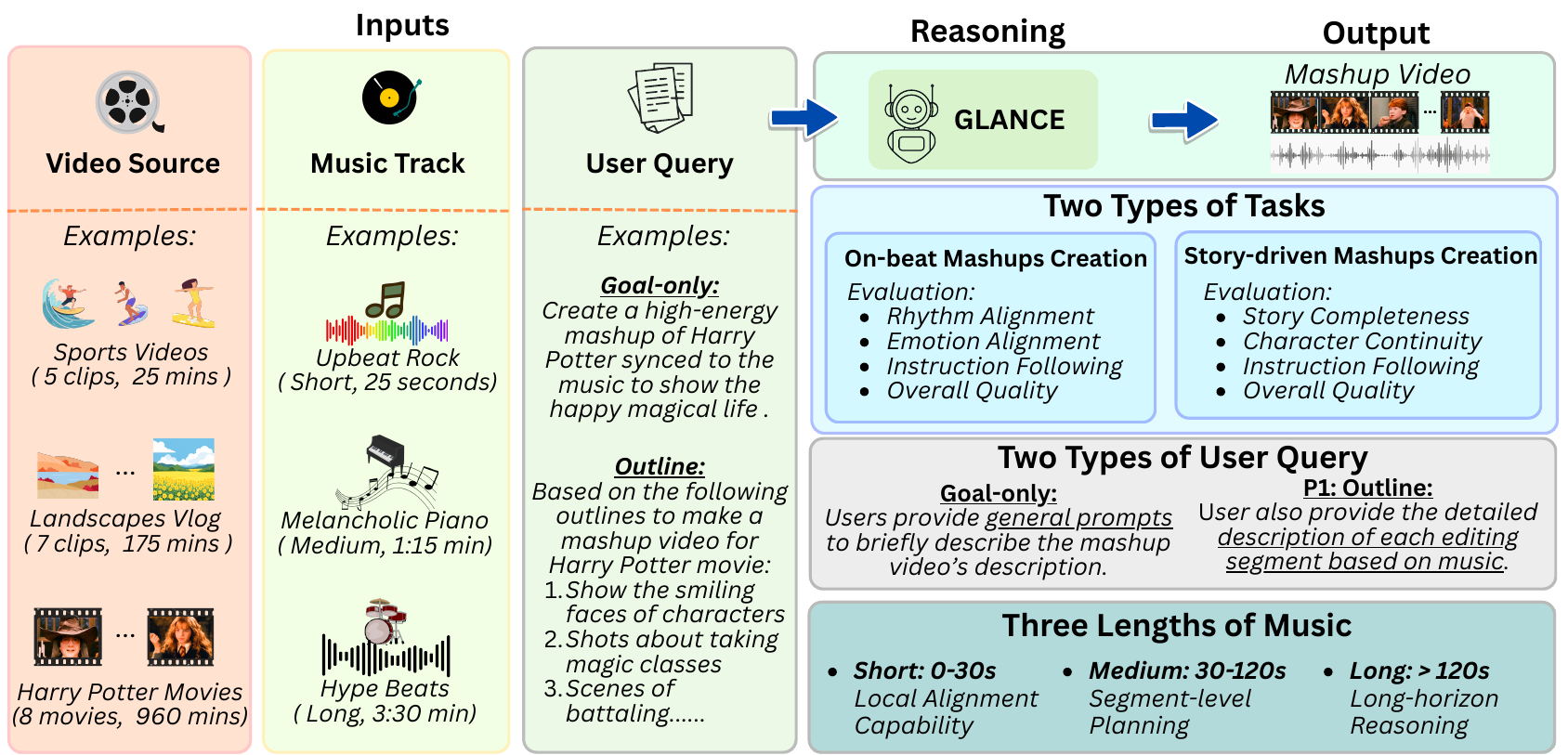}
  \caption{The illustration of MVEBench.}
  \label{fig:mvebench}
\end{teaserfigure}


\maketitle

\section{Introduction}
Video non-linear editing (NLE) aims to construct a new video timeline by selecting, rearranging, and refining visual materials from one or multiple source videos for flexible content reuse, such as retelling stories or expressing emotions. Within this paradigm, \textbf{music-grounded mashup creation} has emerged as an important multimedia scenario, where the synthesized timeline must align with a music track both rhythmically and emotionally. In practice, human editors rely on professional tools such as Adobe Premiere Pro \cite{adobe_premiere}, Final Cut Pro \cite{final_cut_pro}, and DaVinci Resolve \cite{davinci_resolve} to support flexible workflows. However, producing high-quality mashups remains labor-intensive due to large collections of source videos and the iterative process of editing and revision. To reduce human effort, recent commercial products and research efforts have begun to explore multimodal large language models (MLLMs) for automated non-linear video editing. On the product side, platforms such as CapCut \cite{capcut2026} and VEED \cite{veed_io_2026} integrate MLLMs for narration generation, video effects, and music beat analysis. On the research side, several studies investigate multi-agent frameworks for automatic NLE \cite{sandoval2025editduet, zhouvideoagent, xu2024teasergen}. For example, EditDuet \cite{sandoval2025editduet} introduces a perform-and-critic framework that iteratively generates and evaluates candidate edits to produce short videos from large collections of source footage.

Despite these advances, existing approaches still face several critical limitations. 
\textbf{First}, most current multi-agent frameworks rely on predefined pipelines \cite{sandoval2025editduet}, limiting their adaptability to diverse prompts, varying music structures, and heterogeneous source materials. \textbf{Second}, these workflows \cite{xu2024teasergen, zhouvideoagent} do not adequately model the global-local nature of real editing workflows, 
where human editors typically plan globally over the music structure, refine each segment locally, merge subtimelines, and finally revise the composed timeline to resolve cross-segment conflicts. Even well-edited shots can become redundant or inconsistent once assembled. \textbf{Finally}, prior work often evaluates NLE using retrieval accuracy \cite{sandoval2025editduet} or simple success rates \cite{zhouvideoagent}, which are not well suited to open-ended mashup creation. Given the same music track, prompt, and video pool, different editors may produce distinct yet equally valid mashups. For example, a prompt such as \emph{``create a high-energy Harry Potter mashup that conveys a joyful magical life''} may reasonably emphasize either battle scenes or everyday Hogwarts life. This ambiguity makes simple reference-based evaluation inadequate.

These limitations highlight three central research questions (RQ) for music-grounded video NLE:
\textbf{(1) Adaptive Editing:}
How can MLLM-based editing agents dynamically adapt their workflows and tool usage to diverse user intents, music structures, and source materials?
\textbf{(2) Global--Local Coordination:}
How can video editing maintain global coherence while enabling iterative local refinement under long-horizon and cross-segment dependencies?
\textbf{(3) Evaluation:}
How can we design scalable evaluation protocols that capture the multi-dimensional quality of open-ended mashup editing?

To address the first two questions, we propose \textbf{\methodname}, a \emph{global--local coordination multi-agent} framework for music-grounded nonlinear video editing.
To support adaptive editing (\textbf{RQ1}) under diverse prompts, music structures, and source materials, \methodname{} adopts a \emph{bi-loop architecture} that better matches expert editing practice: an outer loop performs adaptive music-aware global planning and task-graph construction, while an inner loop adopts the ``Observe-Think-Act-Verify'' flow to conduct segment-level editing and refinement for each subtask.
This design enables the framework to dynamically organize editing procedures. More importantly, \methodname{} explicitly addresses the \emph{global--local conflict} (\textbf{RQ2}) by introducing a dedicated \emph{global--local coordination mechanism} with both preventive and corrective components.
On the preventive side, a context controller regulates the information exposed to each local editor, including global control signals and the states of previously completed subtasks, so that each segment is optimized with awareness of timeline-level context.
On the corrective side, \methodname{} further resolves the remaining cross-segment inconsistencies through \emph{conflict region decomposition} and \emph{bottom-up dynamic negotiation}.
Together, these designs allow \methodname{} to jointly improve self-adaptivity, local editing quality, and global coherence.

To address the evaluation challenge (\textbf{RQ3}), we construct a new benchmark, \textbf{MVEBench} (Fig. \ref{fig:mvebench}), that factorizes editing difficulty along three orthogonal axes: \emph{task type} (On-Beat vs.\ Story-Driven), \emph{prompt controllability} (from high-level goals to fine-grained instructions), and \emph{music length} (from short clips to long-form settings). This design enables comprehensive and controlled analysis under diverse constraints. The two task types capture complementary creative objectives: \emph{On-Beat} mashup emphasizes rhythm and emotional alignment with the music, while \emph{Story-Driven} mashup focuses on coherent narrative construction with less emphasis on strict beat synchronization. In total, MVEBench contains 319 evaluation samples, covering 645.3 minutes of music and 1,198.2 hours of source video footage. To evaluate this inherently open-ended task, we further introduce an \emph{agent-as-a-judge} evaluation framework that performs scalable and interpretable multi-dimensional assessment.

Experiments on MVEBench demonstrate that \methodname{} consistently outperforms prior research methods and open-source product baselines under the same backbone LLM setting. Using GPT-4o-mini as the backbone, \methodname{} achieves relative improvements of 33.2\% and 15.6\% over the strongest baseline on the two task types, respectively. The improvement is especially evident on more challenging subsets with longer music and less specific prompts, suggesting that \methodname{} is particularly effective when long-horizon planning and coordination are required. Human evaluation further confirms the quality of the generated results, while its high correlation with agent-based evaluation also validates the proposed agent-as-a-judge framework.

The contributions of our work can be summarized as follows:
\begin{itemize}
    \item \textbf{Adaptive editing framework.} We propose \textbf{\methodname{}}, a unified self-adaptive multi-agent framework for music-grounded video NLE that supports dynamic workflow control.

    \item \textbf{Global--local coordination.} We introduce a novel global-local coordination mechanism to mitigate optimization conflicts in preventive and corrective manners.

    \item \textbf{Evaluation framework.} We present a comprehensive evaluation benchmark \textbf{MVEBench} and an \emph{agent-as-a-judge} evaluation framework for 
    scalable, multi-dimensional assessment of music-grounded mashup video editing.
\end{itemize}

\section{Related Work}

\paragraph{Multimodal Agentic Framework.}
Recent multimodal agentic frameworks have been applied to diverse tasks, including image understanding~\cite{kelly2024visiongpt, liu2024llava, lu2025octotools, lu2025scaling}, medical image analysis~\cite{li2024mmedagent, xia2025mmedagent, wang2025medagent, chen2026visual}, image generation and editing~\cite{jiang2026genagent, wang2024genartist, chen2025r2i, wang2025imagent, lin2026mildedit}, and video understanding and generation~\cite{wang2024videoagent, zhi2025videoagent2, yang2025vca, zhang2025deep}. 
Recent work particularly focuses on \emph{long-form video understanding}, where multimodal large language models (MLLMs) act as agents that actively decide what visual evidence to inspect, when to invoke tools, and how to reason over long temporal contexts. 
VideoAgent~\cite{wang2024videoagent} formulates video question answering as an agentic evidence-gathering process. Subsequent studies extend this paradigm through uncertainty-aware reasoning~\cite{zhi2025videoagent2}, curiosity-driven exploration~\cite{yang2025vca}, tool-based search~\cite{zhang2025deep}, motion priors~\cite{liu2025flow4agent}, graph-based reasoning~\cite{chu2025graphvideoagent,shen2025vgent}, streaming anticipation~\cite{yang2025streamagent}, and multi-agent collaboration with reward feedback~\cite{kugo2025videomultiagents,zhou2025reagent,liu2025longvideoagent}. 
In contrast, our work targets \emph{video nonlinear editing}, shifting the focus from answer-centric evidence retrieval to open-ended timeline-level edit generation.

\paragraph{AI-assisted Video Non-linear Editing.}
Prior research on AI-assisted video editing spans multiple directions. Early work focuses on structured scenarios with explicit cinematic rules, such as dialogue-driven editing~\cite{leake2017computational}. Other studies investigate shot-level organization, including cinematography-aware shot assembly~\cite{zhang2025cinematographic}, shot ordering with dedicated benchmarks~\cite{li2025shot}, and narrative-aware editing~\cite{wang2025long}. 
Related work also formulates editing as long-to-short video transformation. Lotus~\cite{barua2025lotus} combines abstractive and extractive summarization, RankCut~\cite{shah2026rankcut} performs transcript-based ranking to select excerpts, TeaserGen~\cite{xu2024teasergen} generates documentary teasers via a narration-centered pipeline, and Repurpose-10K~\cite{wu2025video} introduces a large-scale benchmark for short-form repurposing. 
With the rise of large language models, recent studies explore language-mediated and agent-based editing. EditDuet~\cite{sandoval2025editduet} formulates editing as a multi-agent framework,~\cite{li2025shots} connects shot-level content with editing decisions via language representations, VideoAgent~\cite{zhouvideoagent} moves toward a general agentic framework, and~\cite{zhu2025self} studies iterative refinement for trailer generation. 
However, these methods rely on fixed workflows and do not model the \emph{self-adaptive nature of human expert editing}, nor address the global–local optimization conflict in timeline composition. 

\paragraph{Agentic Framework Algorithms}
Recent work has explored increasingly sophisticated \emph{agentic algorithms} for improving LLM inference-time reasoning and collaboration. DyLAN \cite{liu2024dynamic} dynamically selects a task-specific team of agents and organizes their communication through a query-dependent collaboration structure, enabling adaptive multi-agent problem solving. GPTSwarm \cite{zhuge2024gptswarm} abstracts language agents as optimizable computational graphs and improves both node-level prompting and edge-level orchestration, while Graph-of-Agents \cite{yungraph} further models heterogeneous agents as a relevance-aware graph with node sampling, edge sampling, bidirectional message passing, and graph pooling. Beyond multi-agent collaboration, AB-MCTS \cite{inoue2025wider} studies inference-time search by adaptively deciding whether to expand new branches or deepen existing ones based on external feedback. ToG-2 \cite{ma2024think} interleaves graph retrieval and context retrieval for iterative knowledge-guided reasoning, and recent self-improving agent frameworks \cite{acikgoz2025self} explore uncertainty-aware test-time adaptation through synthetic data generation and temporary fine-tuning. These methods substantially advance generic agentic reasoning, search, and coordination. However, they are primarily designed for tasks such as question answering, coding, or knowledge-intensive reasoning, where the output is a discrete answer, program, or textual solution. In contrast, our work targets \emph{video non-linear editing}, where the agent must construct and revise a temporally coherent edited timeline under coupled global constraints. This requires not only adaptive decomposition and coordination, but also edit-specific global--local optimization over interdependent timeline decisions, which is largely beyond the scope of existing general-purpose agentic algorithms.

\begin{figure*}[t]
    \centering
    \includegraphics[width=0.99\linewidth]{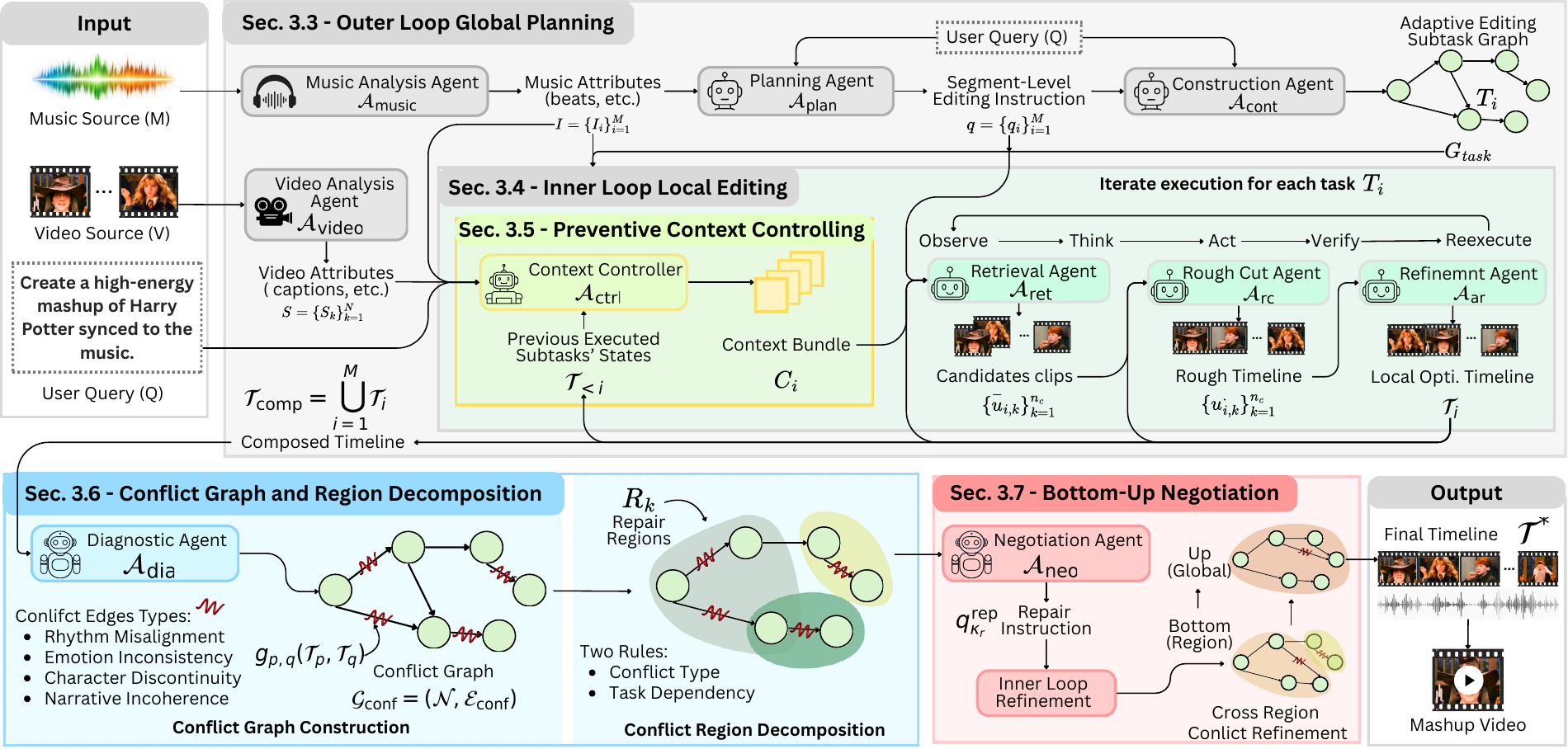}
    \vspace{-2mm}
    \caption{Overview of the \textbf{\methodname{}} framework.}
    \label{fig:glance_framework}
\end{figure*}

\section{Method}
\label{sec:method}

\subsection{Problem Formulation}
\label{subsec:problem_formulation}

Let $\mathcal{Q}$ denote the user editing intent, $\mathcal{M}$ denote the music track, and let $\mathcal{V}=\{v_1,\ldots,v_N\}$ denote the source video collection. The goal of music-guided mashup creation is to generate an edited timeline $\mathcal{T}=\{u_k\}_{k=1}^{K}$, where each timeline unit $u_k$ is a subclip from one of the source videos. The main problem can be formulated as an optimization problem:
\begin{equation}
\mathcal{T}^{*}
=
\arg\max_{\mathcal{T}}
S(\mathcal{T};\mathcal{Q},\mathcal{M},\mathcal{V}),
\label{eq:overall_objective}
\end{equation}
where $S(\cdot)$ is a multi-objective score measuring the quality of the output mashup video, such as story completeness, global emotion alignment, and overall quality.

\subsection{Overview of \methodname{}}
\label{subsec:glance_overview}
As is shown in Figure \ref{fig:glance_framework}, \methodname{} is a multi-agent collaboration framework that consists of two tightly coupled optimization loops: an outer loop (Sec. \ref{subsec:outer_loop}) for global planning and controlling, and an inner loop (Sec. \ref{subsec:inner_loop}) for segment-level editing and refinement. The overall algorithms for outer loop and inner loop are shown in Alg. \ref{alg:outer_loop} and Alg. \ref{alg:inner_loop}, respectively.

Specifically, given a music track $\mathcal{M}$, the outer loop of \methodname{} first analyzes its structure and decomposes it into multiple music intervals. Based on these intervals, the outer loop constructs an adaptive editing task graph, where each node corresponds to a segment-level editing subtask and each edge captures dependency relations among segments (e.g., task 1 must be completed before task 2). For each subtask, the inner loop generates a sub-timeline $\mathcal{T}_i$ by optimizing 
$\mathcal{T}_i^{*}
=
\arg\max_{\mathcal{T}_i}
s_i(\mathcal{T}_i)$,
where $s_i(\mathcal{T}_i)$ measures the quality of segment-level editing. Finally, the outer loop composes all segment-level results to form the final mashup timeline:
$\mathcal{T}
=
\bigcup_{i=1}^{M}\mathcal{T}_i.$

This self-adaptive formulation introduces another challenging optimization problem: highly-optimized local solutions do not necessarily remain optimal after subtimelines composition because of the emerging cross-segment conflicts and global conflicts. As a result, the full-timeline objective is generally non-separable:
\begin{equation}
S_{\text{global}}(\mathcal{T})
=
\sum_{i=1}^{M} s_i(\mathcal{T}_i)
+
\sum_{1\leq i < j \leq M} g_{ij}(\mathcal{T}_i,\mathcal{T}_j)
+
h(\mathcal{T}),
\label{eq:factorized_global_objective}
\end{equation}
where $g_{ij}(\mathcal{T}_i,\mathcal{T}_j)$ models cross-segment compatibility, penalizing issues such as repeated shots, inconsistent character portrayal, or mismatched transitions, and $h(\mathcal{T})$ captures higher-order timeline-level properties. To address this challenge, we novelly propose \emph{global--local coordination strategy} which coordinates outer loop and inner loop in both a \emph{preventive} and a \emph{corrective} manner.
Before local editing, the outer loop provides global control signals and accumulated execution context to guide each inner-loop subtask, reducing conflict-prone local decisions early (Sec.~\ref{subsec:preventive_control}).
After sub-timelines are composed, \methodname{} further identifies and resolves the remaining cross-segment inconsistencies through graph-based conflict region decomposition (Sec.~\ref{subsec:conflict_graph}) and bottom-up dynamic negotiation (Sec.~\ref{subsec:negotiation}).

\subsection{Outer-Loop Global Planning}
\label{subsec:outer_loop}
The outer loop acts as a global controller that organizes the entire editing process before segment-level execution. 
Given a music track $\mathcal{M}$ and a user query $\mathcal{Q}$, \methodname{} first invokes a music analysis agent $\mathcal{A}_{\text{music}}$ to analyze the structure of the music and decompose it into $M$ music segments, $\mathcal{M} = \{m_i\}_{i=1}^{M}$.
For each segment $m_i$, the music analysis agent extracts rhythm and affective cues. 
Formally, each segment is associated with a music attribute vector:
\begin{equation}
\mathcal{I}_i = \mathcal{A}_{\text{music}}(m_i),
\end{equation}
where $\mathcal{I}_i$ consists of the energy profile, emotional attribute, and beat information of the segment. 
The set of segment attributes is denoted as $\mathcal{I}=\{\mathcal{I}_i\}_{i=1}^{M}$. Then, a planning agent $\mathcal{A}_{\text{plan}}$ generates a editing instruction for each music segment:
\begin{equation}
q_i = \mathcal{A}_{\text{plan}}(\mathcal{I}, \mathcal{Q}),
\end{equation}
where $q_i$ specifies the desired editing intent for segment $m_i$, such as semantic theme, emotional tone, or rhythm alignment constraints. After obtaining the segment-level instructions $q=\{q_i\}_{i=1}^{M}$, a construction agent $\mathcal{A}_{\text{cont}}$ organizes these editing tasks into a dependency-aware workflow represented as a directed acyclic graph (DAG):
\begin{equation}
\mathcal{G}_{\text{task}}=(\mathcal{N},\mathcal{E})
=
\mathcal{A}_{\text{con}}(\{q_i\}_{i=1}^{M}),
\end{equation}
where each node $T_i \in \mathcal{N}$ corresponds to an editing task associated with music segment $m_i$, and each directed edge $(T_i,T_j)\in\mathcal{E}$ represents a precedence dependency indicating that task $T_j$ can only be executed after the completion of $T_i$. 

In parallel with music analysis, a video analysis agent $\mathcal{A}_{\text{video}}$ performs high-level understanding of the source video collection $\mathcal{V}=\{v_k\}_{k=1}^{N}$. 
This module extracts structured metadata, including scene boundaries, visual captions, and semantic keywords:
\begin{equation}
\mathcal{S}_k
=
\mathcal{A}_{\text{video}}(v_k),
\end{equation}
where $\mathcal{S}_k$ denotes the semantic representation of video $v_k$. $\mathcal{S}=\{\mathcal{S}_k\}_{k=1}^{N}$ denotes the total video information. These representations provide searchable semantic cues that support subsequent clip retrieval and segment-level editing decisions in the inner loop.

\subsection{Inner-Loop Local Editing}
\label{subsec:inner_loop}

Given the execution graph constructed by the outer loop, the inner loop processes each segment-level editing task sequentially according to the graph order. 
Each subtask corresponds to a specific music interval $m_i$ with local editing instruction $q_i$, music segment information $\mathcal{I}_i$, and overall video information $\mathcal{S}$. From a high-level perspective, the inner loop follows an iterative decision-making cycle:
\begin{equation}
\begin{aligned}
\texttt{Observe}
\rightarrow
\texttt{Think}
\rightarrow
\texttt{Act}
\rightarrow
\texttt{Verify}
\rightarrow
\texttt{Reexecute}.
\end{aligned}
\label{eq:inner_cycle}
\end{equation}

In practice, the inner-loop editing process is realized through three specialized agents responsible for different stages of the pipeline: 
(1) clip retrieval, 
(2) rough-cut construction, and 
(3) alignment and refinement. 
The first two agents mainly implement the ``Observe--Think--Act'' stages, while the refinement agent performs verification and may trigger re-execution of previous stages if inconsistencies are detected. The full decision-making cycle is not applied to every stage, as such a design would be computationally inefficient.

\paragraph{Clip Retrieval.}
The retrieval agent $\mathcal{A}_{\text{ret}}$ first analyzes the local editing instruction $q_i$ and generates multiple retrieval queries in order to capture diverse visual expressions required by the segment. 
Multiple queries are necessary because a single music segment often requires several shots to express a narrative event or emotional progression. Formally, the retrieval agent generates multiple retrieval queries $\{q^{\text{ret}}_{i,t}\}_{t=1}^{n_r}$, such as ``the yong harry potter's smile face showing the happiness'', where $n_r$ denotes the number of generated retrieval queries, and then the agent execute the retrieval process to get video clips for this segment. Formally, we have:
\begin{equation}
\{\hat{u}_{i,k}\}_{k=1}^{n_c} = \mathcal{A}_{\text{ret}}(q_i).
\end{equation}

More details of the clip retrieval agent are discussed in Appendix~\ref{app:innerloop}.

\paragraph{Rough-Cut Generation.}
Given the retrieved clip candidates, the rough-cut agent $\mathcal{A}_{\text{rc}}$ constructs a preliminary segment-level timeline. 
This agent ranks and assembles clips according to multiple criteria, including semantic relevance to input user query and subtask editing instruction, visual quality, and temporal compatibility with the target music segment. Formally, the rough-cut agent produces a preliminary sub-timeline
\begin{equation}
\{u^{*}_{i,k}\}_{k=1}^{n_c}
=
\mathcal{A}_{\text{rc}}
\left(
\{\hat{u}_{i,k}\}_{k=1}^{n_c}, 
q_i,
\mathcal{Q},
\mathcal{S},
\mathcal{I}_i
\right).
\end{equation}

\paragraph{Alignment and Refinement.}
The rough-cut result may still violate several local constraints. 
Typical issues include 
(i) duration mismatch between the assembled clips and the music segment, 
(ii) shot transitions that are not synchronized with music beats, and 
(iii) character, semantic or emotional inconsistency among clips or with the intended segment objective. To address these issues, the alignment and refinement agent $\mathcal{A}_{\text{ar}}$ performs a comprehensive diagnostic over the preliminary sub-timeline. 
This process combines algorithmic signals (e.g., beat detection and duration constraints) with LLM-based reasoning to identify potential inconsistencies. Formally, the refinement agent produces the final segment-level result:
\begin{equation}
\mathcal{T}_i=\{u_{i,k}\}_{k=1}^{n_c}
=
\mathcal{A}_{\text{ar}}
\left(
\{u^{*}_{i,k}\}_{k=1}^{n_c}
\right).
\end{equation} 

If the refinement agent detects structural issues that cannot be resolved locally, it may trigger a re-execution of earlier stages. 

After all subtasks in the task graph finish execution, the \emph{outer loop} again composes all subtimeline outputs into the final result global mashup video timeline:
\begin{equation}
    \mathcal{T}_{\text{comp}} = \bigcup_{i=1}^{M}\mathcal{T}_i.
\end{equation}

\subsection{Preventive Context Controlling}
\label{subsec:preventive_control}

Although each inner-loop editor optimizes its local objective, the final merged timeline may not be globally optimized because of the emerging cross-segment and global conflicts, as is shown in Eq. \ref{eq:factorized_global_objective}. To reduce such conflict-prone behaviors early, \methodname{} introduces a \emph{preventive context control mechanism}. 
Specifically, controller agent $\mathcal{A}_{\text{ctrl}}$ selectively determines the information that should be visible to the current subtask. Formally, for the editing task associated with music segment $m_i$, the controller constructs a context bundle
\begin{equation}
\mathcal{C}_i
=
\mathcal{A}_{\text{ctrl}}
\left(
\mathcal{I}_i,
q_i,
\mathcal{T}_{<i},
\mathcal{S}
\right),
\end{equation}
where $\mathcal{T}_{<i}$ represents the set of previously completed subtimelines. The controller filters and organizes these signals to construct a task-specific observation space for the inner-loop agents. 
In practice, the context bundle typically contains three components: 
(1) local editing objectives derived from the music segment and user intent; 
(2) summarized execution states of previously completed segments; and 
(3) a constrained retrieval scope over the video metadata. The inner-loop editing agents then operate conditioned on this context:
\begin{equation}
\mathcal{T}_i
=
\mathcal{A}_{\text{inner}}
\left(
\mathcal{C}_i
\right),
\end{equation}
which ensures that local editing decisions remain consistent with both the global plan and previously generated timeline segments. 

The reason that we avoid exposing the full states of previously completed tasks is that, in practice, the accumulated state history can become excessively long and may interfere with the preventive behavior. This preventive design reduces the likelihood of repeated footage, incompatible transitions, or narrative discontinuities during early-stage editing. Nevertheless, since some cross-segment dependencies cannot be fully anticipated beforehand, residual conflicts may still emerge after the subtimelines are composed. To address these remaining inconsistencies, \methodname{} further introduces corrective coordination mechanisms based on conflict graph decomposition and bottom-up negotiation.

\subsection{Conflict Graph and Regional Decomposition}
\label{subsec:conflict_graph}
After the composition of the final timeline output, \methodname{} constructs a \emph{conflict graph} over the current composed timeline. Specifically, given the current segment timelines $\{\mathcal{T}_i\}_{i=1}^{M}$ and the task graph $\mathcal{G}$, a diagnostic agent $\mathcal{A}_{\text{dia}}$ analyzes each adjacent segment pair $(i,j)\in\mathcal{E}$.
The agent first invokes perception tools (e.g., MLLMs) to extract semantic signals such as emotion labels, character identities, and captions, and then performs reasoning to detect cross-segment inconsistencies. Formally, the conflict graph is defined as
\begin{equation}
\mathcal{G}_{\text{conf}}
=
(\mathcal{N},\mathcal{E}_{\text{conf}})
=
\mathcal{A}_{\text{dia}}(\{\mathcal{T}_i\}),
\end{equation}
where each node
$n_i =
(\mathcal{T}_i, m_i, q_i, \texttt{TimelineMem}_i)$
represents a segment-level editing result with its execution context. Intermediate reasoning states are stored in $\texttt{TimelineMem}_i$. An edge $(n_p,n_q)\in\mathcal{E}_{\text{conf}}$ indicates that the two segments jointly violate one or some constraints. We define a conflict predicate as
\begin{equation}
\begin{aligned}
g_{p,q}(\mathcal{T}_p,\mathcal{T}_q)
=
\mathbb{I}\!\big[
&\phi_{\text{rhythm}}(n_p,n_q)
\vee
\phi_{\text{emotion}}(n_p,n_q) \\
&\vee
\phi_{\text{character}}(n_p,n_q)
\vee
\phi_{\text{story}}(n_p,n_q)
\big],
\end{aligned}
\label{eq:conflict_predicate}
\end{equation}
where the predicates test rhythm misalignment, emotion inconsistency, character discontinuity, and narrative incoherence.

\paragraph{Conflict-aware regional decomposition.}
A straightforward repair strategy would resolve conflicts edge-by-edge.
However, pairwise repairs can be unstable: fixing a conflict between $(i,j)$ can introduce a new conflict with $(j,k)$, and resolving $(j,k)$ can invalidate the previous fix for $(i,j)$.
Alternatively, repairing entire connected components of $\mathcal{G}_{\text{conf}}$ may degenerate into global re-optimization when the graph becomes dense, which requires expensive computational cost due to the lengthy context information. To balance stability and efficiency, \methodname{} performs \emph{conflict-aware regional decomposition}.
Each repair region is initialized from a detected conflict edge $(n_p,n_q)\in\mathcal{E}_{\text{conf}}$ and expanded to include structurally related nodes according to two rules:
(i) shared the same conflict types, and
(ii) dependency relations in the editing task graph.
Formally, a repair region is defined as
\begin{equation}
\mathcal{R}_k
=\{\kappa_r\}_{r=1}^{R}=
\texttt{Expand}(n_p,n_q,\mathcal{G}),
\end{equation}
which induces a local subgraph
\begin{equation}
\mathcal{G}_k=(\mathcal{N}_k,\mathcal{E}_k),\quad \mathcal{N}_k\subseteq\mathcal{N}.
\end{equation}

In our experiments, we further impose an upper bound on the size of each conflict region, limiting the number of nodes to 
\begin{equation}
\min(4, |\mathcal{N}|/4).
\end{equation}

Please note that one node can be decomposed into more than one region due to different conflict types. The efficiency of this mechanism is analyzed in Appendix~\ref{app:effi-conflict-region-decomp}.

\subsection{Bottom-Up Conflict-Aware Negotiation}
\label{subsec:negotiation}

Given the detected conflict regions $\mathcal{R}_k$, \methodname{} performs a bottom-up negotiation procedure to revise the current timeline.
Rather than re-optimizing the entire timeline globally, the agent resolves conflicts progressively from local regions toward the global composition. For each conflict region $\kappa_r$, the negotiation agent $\mathcal{A}{\text{neo}}$ generates repair instructions $q_{\kappa_r}^{\text{rep}}$. The inner-loop editing procedure is then applied to each instruction, modifying only the segments within the region while keeping the remainder of the timeline fixed. After regional repairs are applied, the diagnostic agent $\mathcal{A}_{\text{dia}}$ re-evaluates compatibility across region boundaries. If new cross-region conflicts emerge, the corresponding regions are merged into a larger region, and the negotiation and refinement steps are repeated. Finally, a global refinement step is performed once no further pairwise conflicts remain. The stop criterion is reaching the predefined number (40 in our experiments) or reaching the final global refinement. Through this bottom-up negotiation mechanism, \methodname{} progressively propagates local corrections across neighboring regions and global optimization.

\begin{table*}[t]
\centering
\small
\caption{Overall comparison on the full MVEBench.
Judge-based metrics are reported on a 1--5 scale.
Best results are in \textbf{bold}; second-best are underlined. 
All baselines utilize GPT-4o-mini as the backbone model, except for specific demonstrations for \methodname{}. The results are the average score of all types of configuration as described in Section \ref{sec:mvebench}.} 
\vspace{-2mm}
\label{tab:main_overall}

\begin{tabular}{l|cccc|cccc}
\toprule
& \multicolumn{4}{c}{\textbf{On-beat Video}} 
& \multicolumn{4}{c}{\textbf{Story-driven Video}} \\
\cmidrule(lr){2-5} \cmidrule(lr){6-9}
Method
& \makecell{Rhythm\\Alignment$\uparrow$}
& \makecell{Emotion\\Alignment$\uparrow$}
& \makecell{Instruction\\Following$\uparrow$}
& \makecell{Overall\\Quality$\uparrow$}
& \makecell{Story\\Completeness$\uparrow$}
& \makecell{Character\\Continuity$\uparrow$}
& \makecell{Instruction\\Following$\uparrow$}
& \makecell{Overall\\Quality$\uparrow$} \\
\midrule
\rowcolor{gray!8}\multicolumn{9}{l}{\textit{\textbf{LLM-based Agentic Framework}}} \\ 
\midrule
CoT                    & 2.06 & 1.95 & 1.98 & 1.47 & 2.12 & 2.05 & 2.04 & 1.65 \\
GPTSwarm               & 2.31 & 2.22 & 2.24 & 1.64 & 2.34 & 2.29 & 2.28 & 1.95 \\

\midrule
\rowcolor{gray!8}\multicolumn{9}{l}{\textit{\textbf{Multi-modal Video Editing Agentic Framework}}} \\ 
\midrule
TeaserGen              & 2.01 & 1.90 & 1.94 & 1.76 & 2.43 & 2.33 & 2.34 & 2.10 \\
EditDuet               & 2.85 & 2.73 & 2.76 & 2.21 & 2.94 & 2.82 & 2.84 & 2.81 \\
VideoAgent             & 3.04 & 2.94 & 2.97 & 2.53 & 3.13 & 3.04 & 3.06 & 2.88 \\

\midrule
\rowcolor{gray!8}\multicolumn{9}{l}{\textit{\textbf{Video Editing Agentic-based Product}}} \\ 
\midrule
FunCLIP                & 2.63 & 2.56 & 2.59 & 2.23 & 3.02 & 2.95 & 2.97 & 2.59 \\
NarratoAI              & 2.69 & 2.61 & 2.64 & 2.24 & 3.08 & 3.01 & 3.02 & 2.65 \\

\midrule
\rowcolor{gray!8}\multicolumn{9}{l}{\textit{\textbf{\methodname{} (Ours)}}} \\ 
\midrule
\methodname{} (Qwen3-VL-8B)   & 2.86 & 2.74 & 2.77 & 2.69 & 2.93 & 2.84 & 2.86 & 2.68 \\
\methodname{} (Qwen3-VL-30B)  & 3.16 & 3.03 & 3.05 & 2.88 & 3.24 & 3.14 & 3.16 & 3.01 \\
\methodname{} (Gemini-2-pro)        & \textbf{3.72} & \underline{3.54} & \textbf{3.66} & \textbf{3.45} & \textbf{3.75} & \textbf{3.68} & \textbf{3.95} & \textbf{3.42} \\
\methodname{} (GPT-4o-mini)           & \underline{3.61} & \textbf{3.65} & \underline{3.58} & \underline{3.37} & \underline{3.69} & \underline{3.59} & \underline{3.87} & \underline{3.33} \\

\bottomrule
\end{tabular}
\end{table*}

\section{MVEBench}
\label{sec:mvebench}

To comprehensively evaluate music-guided mashup video editing methods, we introduce \textbf{MVEBench}, a \textbf{M}usic-guided \textbf{V}ideo \textbf{E}diting Benchmark designed for evaluating agent-based non-linear video editing frameworks. Each instance in MVEBench consists of three components
$(\mathcal{Q}, \mathcal{V}, \mathcal{M})$,
where $\mathcal{Q}$ denotes the user editing intent, $\mathcal{V}$ represents a set of source videos, and $\mathcal{M}$ is the music track guiding the editing process. The benchmark is designed to reflect realistic editing scenarios encountered in music-driven mashup video creation.

\paragraph{Data Collection Pipeline.}
We first collect professionally created mashup videos from video-sharing platforms such as Bilibili \cite{bilibili} and YouTube \cite{youtube} using keywords including \emph{mashup video}, \emph{video remix}, and \emph{on-beat editing}. For each collected mashup video, annotators identify the corresponding music track $\mathcal{M}$ and the set of source videos $\mathcal{V}$. The original source materials are then retrieved from public or license-free repositories to ensure legal redistribution. Given the collected music and source videos, human annotators write a natural-language description representing the \emph{user editing intent} $\mathcal{Q}$. Each intent is verified through automatic checking by a large language model and cross-validation by independent annotators. Notably, the collected mashup videos are \textbf{not treated as ground-truth outputs}. Since mashup creation is inherently open-ended, multiple valid editing results may exist for the same intent.

\paragraph{Benchmark Taxonomy.} MVEBench organizes editing tasks along several dimensions for a comprehensive evaluation. First, tasks are divided into two editing styles: \emph{on-beat mashups} (\textbf{O}), emphasizing rhythmic synchronization with music, and \emph{story-driven mashups} (\textbf{S}), focusing on narrative coherence and semantic progression. Second, music tracks are categorized into \textbf{Sh} (short), \textbf{Me} (medium), and \textbf{Lo} (long), where longer music generally requires stronger long-horizon planning. Third, prompts are grouped into \textbf{GP} (general prompts describing high-level goals) and \textbf{DP} (detailed prompts providing structured editing guidance). Detailed definitions are provided in Appendix~\ref{app:details-of-taxonomy}. The benchmark combines the three aspects to form a total of eight types of tasks. Several configuration combinations are intentionally excluded when constructing MVEBench. First, \textbf{Onbeat–LongMusic–GeneralPrompt} is excluded. When the music duration is long while only a very general prompt is provided, the editing objective becomes underspecified, making the resulting videos difficult to evaluate in a consistent and reliable manner. Second, \textbf{Onbeat–ShortMusic–DetailedPrompt} is excluded. Short music clips do not provide sufficient temporal capacity to accommodate detailed prompt specifications. In practice, such configurations make it difficult to meaningfully evaluate the instruction-following ability of editing agents. Third, all \textbf{StoryDriven–ShortMusic} configurations are excluded regardless of the prompt type. Story-driven editing requires sufficient temporal duration to convey narrative progression, which short music clips cannot adequately support. For these reasons, the above combinations are excluded from the benchmark design. We report the final statistics of MVEBench in Table~\ref{tab:mvebench_stats}.

\begin{table}[t]
\centering
\small
\caption{Statistics of MVEBench under different task configurations. The unit of music length is second (s), while the unit of video source length is hour (h). ``\#. Samples'' represent the number of tasks for each configuration.}
\label{tab:mvebench_stats}

\begin{tabular}{lccc}
\toprule
Config & \#. Samples & Avg. Music Len. & Avg. Source Video Len. \\
\midrule

O-Sh-GP & 45 & 27.9s & 1.7h \\
O-Me-GP & 35 & 83.7s & 3.1h \\
O-Me-DP & 35 & 83.7s & 3.1h \\
O-Lo-DP & 44 & 154.2s & 5.5h \\

S-Me-GP & 37 & 85.6s & 2.9h \\
S-Lo-GP & 45 & 182.1s & 5.2h \\
S-Me-DP & 34 & 85.1s & 2.9h \\
S-Lo-DP & 44 & 179.8s & 5.2h \\

\bottomrule
\end{tabular}

\end{table}

\section{Agentic Evaluation Framework}
\label{subsec:agentic_eval}

To address the open-ended evaluation challenge, we propose an \textbf{agent-as-a-judge} evaluation framework that formulates mashup assessment as a structured multi-stage reasoning process. 
Given the edited timeline $\mathcal{T}$, music track $\mathcal{M}$, and user intent $\mathcal{Q}$, the framework first invokes a set of perception agents to extract grounded evidence from music and video. 
Formally, a set of analysis agents
$
\{\mathcal{A}_{\text{music}}, \mathcal{A}_{\text{video}}\}$
produce structured observations, such as music beat information and detailed captions of videos.
A reasoning agent $\mathcal{A}_{\text{judge}}$ then aggregates these observations and produces dimension-specific evaluation scores: $\mathbf{s} = \mathcal{A}_{\text{judge}}(\mathcal{E}, \mathcal{Q})$,
where $\mathbf{s}$ contains the scores for different evaluation dimensions. We introduce more details, motivations, and discussions of the agent-as-a-judge framework in Appendix~\ref{app:agent-evaluation-framework}.

\paragraph{Evaluation Dimensions.}
We design separate evaluation protocols for the two task families in MVEBench.
For \textbf{on-beat mashup video creation}, we evaluate \emph{Rhythm Alignment}, \emph{Emotion Alignment}, \emph{Instruction Following}, and \emph{Overall Quality}.
These videos primarily emphasize synchronization with music and affective consistency, rather than complete story development.
For \textbf{story-driven mashup video creation}, we evaluate \emph{Story Completeness}, \emph{Character Continuity}, \emph{Instruction Following}, and \emph{Overall Quality}.
In this setting, the key challenge is whether the edited video forms a coherent and understandable story while maintaining consistent characters and globally reasonable composition. The details of each evaluation criterion and the implementations are shown in Appendix~\ref{app:details-evaluation-criteria}.

\begin{table*}[t]
\centering
\small
\caption{Ablation of global--local coordination modules on \textbf{on-beat} mashup video creation. Efficiency is reported as normalized processing efficiency measured by token usage, with higher being better.}
\vspace{-2mm}
\label{tab:ablation_components_onbeat}
\begin{tabular}{lcccccccc}
\toprule
Variant
& Preventive
& Region Decomp.
& Bottom-up Negot.
& \makecell{Rhythm\\Alignment$\uparrow$}
& \makecell{Emotion\\Alignment$\uparrow$}
& \makecell{Instruction\\Following$\uparrow$}
& \makecell{Overall\\Quality$\uparrow$}
& \makecell{Efficiency$\uparrow$} \\
\midrule
Only Preventive Controller & \checkmark &  &  & 3.18 & 3.07 & 3.12 & 3.05 & \textbf{1.18} \\
w/o Bottom-up Negotiation  & \checkmark & \checkmark &  & 3.36 & 3.25 & 3.30 & 3.24 & 1.10 \\
w/o Region Decomposition   & \checkmark &  & \checkmark & 3.46 & 3.35 & 3.39 & 3.33 & 0.76 \\
w/o Preventive Controller  &  & \checkmark & \checkmark & 3.58 & 3.46 & 3.52 & 3.40 & 0.96 \\
Full \methodname{}                & \checkmark & \checkmark & \checkmark & \textbf{3.72} & \textbf{3.54} & \textbf{3.66} & \textbf{3.45} & 1.00 \\
\bottomrule
\end{tabular}
\end{table*}

\section{Experiments}
\label{sec:experiments}

\subsection{Experimental Settings}
\label{subsec:exp_protocol}
We conduct all experiments on \textbf{MVEBench}, our benchmark for music-guided non-linear video editing.
As introduced in Sec.~\ref{sec:mvebench}, MVEBench covers two task families:
\emph{on-beat mashup} and \emph{story-driven mashup},
with different difficulty levels induced by music duration and planning horizon.
We further consider two levels of user intent, namely \emph{general-level intent} and \emph{structured intent}, in order to evaluate both open-ended planning and controllable instruction following.
Unless otherwise specified, all reported results are averaged over the full test split.

We compare \methodname{} against three groups of representative baselines: (1)\textbf{LLM-based agentic frameworks}: 
\textbf{CoT}~\cite{wei2022chain} is a single-agent prompting paradigm that improves reasoning by eliciting intermediate natural-language reasoning steps before producing the final output. We adapt it as a strong single-agent planning baseline for video editing. \textbf{GPTSwarm}~\cite{zhuge2024gptswarm} models language agents as an optimizable computational graph, where nodes represent agent operations and edges represent information flow, and further improves the framework through node-level prompt optimization and edge-level orchestration optimization. (2) \textbf{Multimodal video editing and video-agent baselines}:
\textbf{TeaserGen}~\cite{xu2024teasergen} is a narration-centered two-stage framework for long-documentary teaser generation: it first generates teaser narration from documentary transcripts using an LLM, and then retrieves or aligns visual content to match the generated narration. \textbf{EditDuet}~\cite{sandoval2025editduet} is a multi-agent non-linear editing framework that frames video editing as a sequential decision-making problem and iteratively refines the timeline through interaction between an \emph{Editor} agent and a \emph{Critic} agent. \textbf{VideoAgent}~\cite{zhouvideoagent} is an all-in-one agentic framework for video understanding and editing that combines shot planning agents, cross-modal retrieval, and self-reflective orchestration over a large pool of specialized editing agents. (3) \textbf{Engineering and product-style baselines}:
We further compare against representative practical editing systems:
\textbf{FunCLIP}~\cite{modelscope_funclip_2025}, and
\textbf{NarratoAI}~\cite{linyqh_narratoai_2026}.
These systems are included to reflect practical retrieval-and-composition pipelines and off-the-shelf editing workflows.


\methodname{} can be instantiated with different reasoning backbones.
In our experiments, we evaluate four variants, including two closed-source models: GPT-4o-mini \cite{openai_gpt_models} and Gemini-2-pro \cite{google_gemini_models}, and two open-source models Qwen3-VL-8B \cite{bai2025qwen3} and Qwen3-VL-30B \cite{bai2025qwen3}. 

\subsection{Main Results}
\label{subsec:main_results}

\subsubsection{Overall benchmark comparison}
Table~\ref{tab:main_overall} shows that \methodname{} consistently achieves the strongest performance across both \emph{on-beat} and \emph{story-driven} settings. Among all baselines, \methodname{} instantiated with closed-source backbones attains the best results on nearly all evaluation criteria, demonstrating the effectiveness of our framework for music-guided mashup video creation. Specifically, \methodname{} (GPT-4o-mini) achieves 3.37 and 3.33 in Overall Quality on the on-beat and story-driven subsets, respectively, outperforming the strongest baseline, \textit{VideoAgent}, by 33.2\% and 15.6\%. Overall Quality scores are generally lower than other metrics because they involve more subjective aesthetic judgments, such as shot composition, transition smoothness, and narrative expressiveness, which are harder to optimize automatically than objective metrics like rhythm alignment or instruction following. Importantly, the gains of \methodname{} are not limited to proprietary models. With the open-source backbone Qwen3-VL-30B, \methodname{} still surpasses \textit{VideoAgent}, achieving 2.88 and 3.01 in Overall Quality for the two settings, suggesting that the improvements mainly stem from the proposed framework rather than the backbone alone. Even with a smaller model, Qwen3-VL-8B, \methodname{} remains competitive with stronger LLM-based agent frameworks and commercial editing products such as GPTSwarm and NarratoAI, indicating good generalization across foundation models of different scales.

\subsubsection{Results by different task type}
The improvement is particularly pronounced on the \emph{story-driven} subset, where long-horizon planning and cross-segment consistency are more critical. \methodname{} (Gemini) achieves 3.75 in Story Completeness, 3.68 in Character Continuity, and 3.95 in Instruction Following, substantially surpassing the strongest baseline, \textit{VideoAgent}, which obtains 3.13, 3.04, and 3.06 on the same metrics. By comparison, although the gains on the \emph{on-beat} subset remain clear, they are relatively smaller: \methodname{} reaches 3.72 in Rhythm Alignment and 3.45 in Overall Quality, compared with 3.04 and 2.53 for \textit{VideoAgent}. This pattern suggests that the main advantage of \methodname{} lies not only in improving local beat-level alignment, but more importantly in coordinating temporally coupled decisions across segments. Such capability is especially important for story-driven video creation, where final quality depends on maintaining narrative progression, character continuity, and global coherence over long temporal horizons.

\begin{figure}[t]
    \centering
\includegraphics[width=0.95\linewidth]{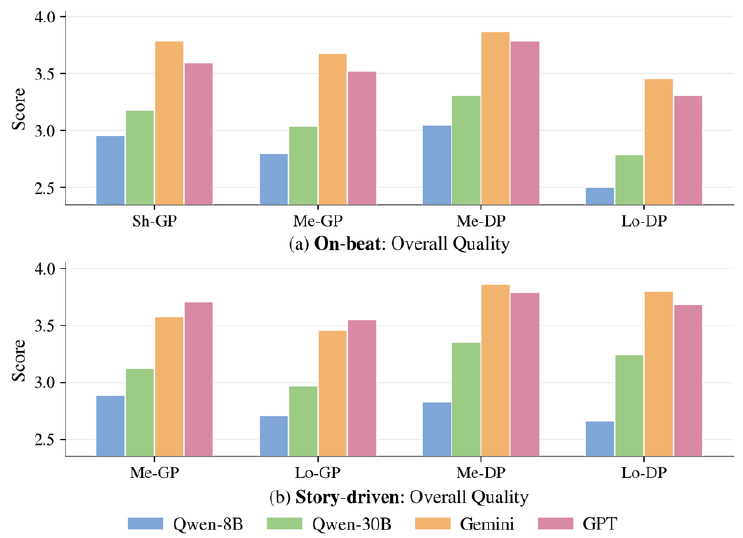}
\vspace{-4mm}
    \caption{Overall quality across different task configurations.}
    %
    \label{fig:difficulty-results-main}
\end{figure}

\subsubsection{Results by difficulty and intent level}
Figure~\ref{fig:difficulty-results-main} reports subset-level results under different music lengths and prompt controllability settings.
Overall, all \methodname{} variants show consistent trends across subsets, while stronger backbones remain more robust as task difficulty increases. For the on-beat setting, the long-music detailed-prompt subset (\textbf{O-Lo-DP}) is consistently the most challenging, whereas medium-length music with detailed prompts (\textbf{O-Me-DP}) and short music with general prompts (\textbf{O-Sh-GP}) are comparatively easier.
For example, in overall quality, \methodname{} (GPT) achieves 3.79 on \textbf{O-Me-DP} but drops to 3.31 on \textbf{O-Lo-DP}, while \methodname{} (Gemini) shows a similar decrease from 3.87 to 3.46.
This indicates that long-horizon on-beat editing remains difficult even for strong backbones due to extended temporal rhythm constraints. A similar pattern appears in the story-driven setting.
Subsets with more structured intent benefit stronger models, while weaker backbones degrade more noticeably under complex narrative requirements.
Meanwhile, general-prompt subsets, especially \textbf{S-Lo-GP}, remain harder than medium-length cases, suggesting that long-form story composition without explicit guidance is particularly challenging. Overall, the advantage of \methodname{} becomes clearer on harder subsets with longer temporal scope or stronger control requirements, highlighting its ability to coordinate multiple interdependent editing decisions under global constraints.

\begin{table}[t]
\centering
\small
\caption{Human evaluation results on sampled On-Beat and Story-Driven subsets.}
\vspace{-2mm}
\label{tab:human_main}
\begin{tabular}{l l c c c}
\toprule
Task Type & Metric & Gemini & Qwen-30B & EditDuet  \\
\midrule
On-Beat & Rhythm Alignment      & 3.43 & 2.91 & 2.57 \\
On-Beat & Emotion Alignment     & 3.31 & 2.78 & 2.45 \\
On-Beat & Instruction Following & 3.38 & 2.83 & 2.48 \\
On-Beat & Overall Quality       & 3.11 & 2.54 & 2.29 \\
Story-Dri. & Story Completeness    & 3.49 & 2.97 & 2.68 \\
Story-Dri. & Character Continuity  & 3.41 & 2.88 & 2.53 \\
Story-Dri. & Instruction Following & 3.45 & 2.93 & 2.60 \\
Story-Dri. & Overall Quality       & 3.26 & 2.62 & 2.29 \\
\bottomrule
\end{tabular}
\end{table}

\subsubsection{Ablation studies on global--local coordination modules}
\label{subsubsec:ablation_coordination}
We conduct ablation studies on \textsc{\methodname{}} with the Gemini backbone to analyze the contribution of the proposed global--local coordination design.
We evaluate four variants:
(1) \emph{w/o preventive context controller}, where local editors do not receive controlled global context or finalized states;
(2) \emph{w/o conflict region decomposition}, where conflicts are resolved directly on connected node pairs instead of grouped regions;
(3) \emph{w/o bottom-up iterative negotiation}, where conflicts are repaired sequentially from left to right on the conflict graph;
and (4) \emph{only preventive context controller}, which keeps the preventive mechanism but removes the corrective modules. Tables~\ref{tab:ablation_components_onbeat} and~\ref{tab:ablation_components_story} show that all modules contribute positively but play different roles.
Removing the preventive controller causes the smallest drop, suggesting that later corrective stages can partially compensate for its absence.
In contrast, using the preventive controller alone performs worst, indicating that preventive control alone is insufficient.
Among the corrective modules, bottom-up negotiation is more critical than conflict region decomposition: removing negotiation leads to larger performance drops (e.g., Rhythm Alignment $3.72 \rightarrow 3.36$), as sequential repair often introduces new cross-segment inconsistencies.
Removing region decomposition results in smaller quality degradation ($3.72 \rightarrow 3.46$) but the lowest efficiency (0.76), since \methodname{} must repeatedly resolve entangled pairwise conflicts.
Overall, the best performance is achieved only when preventive control and corrective coordination are combined, validating the necessity of the full global--local coordination design.

\subsection{Human Evaluation and Judge Consistency}
\label{subsec:human_eval}

\paragraph{Human evaluation.} To validate the proposed agent-as-a-judge protocol, we randomly sample 50 examples from the on-beat task and the story-driven task separately, and invite two human experts with at least three years of mashup editing experience to provide independent ratings for each output. As shown in Table~\ref{tab:human_main}, the ranking under human evaluation is highly consistent with that of the automatic benchmark, and \methodname{} remains the strongest method overall. In the On-beat setting, \methodname{}(Gemini) achieves 3.43 in Rhythm Alignment and 3.11 in Overall Quality, compared with 2.91 and 2.54 for \methodname{}(Qwen3-VL-30B), and 2.57 and 2.29 for EditDuet, respectively.
A similar trend is observed in the Story-Driven setting. These results further suggest that the advantage of \methodname{} is perceptually meaningful. 

\paragraph{Agreement with human evaluation.}
Table~\ref{tab:judge_human_corr} reports the agreement between the judge and human ratings.
We observe strong consistency across all dimensions, indicating that the judge provides a reliable proxy for large-scale benchmarking.
Notably, rhythm alignment and overall quality tend to exhibit the strongest agreement, while story completeness is somewhat more challenging due to its higher subjectivity.

\begin{table}[t]
\centering
\small
\caption{Agreement between the proposed judge and human evaluation.}
\label{tab:judge_human_corr}
\begin{tabular}{l l c c}
\toprule
Task Type & Metric & Spearman $\rho \uparrow$ & Kendall $\tau \uparrow$ \\
\midrule
On-Beat & Rhythm Alignment         & 0.64 & 0.48 \\
On-Beat & Emotion Alignment        & 0.62 & 0.46 \\
On-Beat & Instruction Following    & 0.60 & 0.47 \\
On-Beat & Overall Quality          & 0.66 & 0.50 \\
Story-Driven & Story Completeness   & 0.71 & 0.55 \\
Story-Driven & Character Continuity & 0.68 & 0.52 \\
Story-Driven & Instruction Following    & 0.67 & 0.52 \\
Story-Driven & Overall Quality      & 0.73 & 0.57 \\
\midrule
Average & -- & 0.67 & 0.51 \\
\bottomrule
\end{tabular}
\end{table}

\section{Conclusion}
In this paper, we present \textbf{\methodname{}}, a global--local coordination multi-agent framework for music-grounded mashup video creation, which consists of a bi-loop design that separates global planning from local editing refinement. To address the fundamental global--local optimization conflict arising from subtimeline decomposition, we further propose a coordination mechanism that combines preventive context control with corrective conflict resolution. In addition, we introduce \textbf{MVEBench} and an agent-as-a-judge evaluation framework to enable scalable assessment of mashup video quality. Extensive experiments demonstrate that \methodname{} consistently outperforms existing baselines across diverse editing settings, while ablation studies validate the importance of each proposed component.

\newpage
\bibliographystyle{ACM-Reference-Format}
\bibliography{sample-base}

\appendix

\section{Implementation Details and Qualitative Analysis}
\label{subsec:qualitative}
We will show the detailed implementations including all prompts for each agent, hyperparameters, and codes of our framework, in the supplementary materials. We will also show the real output as qualitative analysis in the supplementary materials.

\section{\methodname{}}

\subsection{The algorithms details}
The details of the inner loop and outer loop's algorithm are shown in Alg. \ref{alg:inner_loop} and \ref{alg:outer_loop}, respectively.

\subsection{Inner Loop}
\label{app:innerloop}
In the clip retrieval step, for each query, the agent adaptively selects appropriate retrieval tools, including vision-language models (e.g., CLIP-based retrieval), multimodal LLM reasoning, or text-based search over pre-generated video descriptions $\mathcal{S}$. 
When necessary, temporal grounding is applied to localize precise time spans within candidate scenes.
\subsection{The efficiency of conflict region decomposition}
\label{app:effi-conflict-region-decomp}
This decomposition significantly reduces computational complexity.
Let $K$ denote the number of candidate solutions per segment.
Global optimization requires $\mathcal{O}(K^{M})$ search over all segments, while regional optimization reduces the cost to $\sum_k \mathcal{O}(K^{|\mathcal{R}_k|})$, where typically $|\mathcal{R}_k|\ll M$.
Therefore, \methodname{} can efficiently resolve coupled conflicts while preserving most previously optimized local results.

\begin{algorithm}[t]
\caption{\methodname{} Outer-Loop Global Planning and Coordination}
\label{alg:outer_loop}
\begin{algorithmic}[1]
\REQUIRE User query $\mathcal{Q}$, music track $\mathcal{M}$, source videos $\mathcal{V}$
\ENSURE Final mashup timeline $\mathcal{T}^{*}$

\STATE $\{m_i\}_{i=1}^{M} \gets \mathcal{A}_{\text{music}}(\mathcal{M})$ \COMMENT{Music segmentation}

\FOR{$i=1$ to $M$}
    \STATE $\mathcal{I}_i \gets \mathcal{A}_{\text{music}}(m_i)$
    \STATE $q_i \gets \mathcal{A}_{\text{plan}}(\mathcal{I},\mathcal{Q})$
\ENDFOR

\STATE $\mathcal{G}_{\text{task}}=(\mathcal{N},\mathcal{E})
\gets
\mathcal{A}_{\text{con}}(\{q_i\}_{i=1}^{M})$ \COMMENT{Build editing DAG}

\FOR{$k=1$ to $N$}
    \STATE $\mathcal{S}_k \gets \mathcal{A}_{\text{video}}(v_k)$
\ENDFOR
\STATE $\mathcal{S} \gets \{\mathcal{S}_k\}_{k=1}^{N}$

\STATE $\mathcal{B} \gets \texttt{TopologicalOrder}(\mathcal{G}_{\text{task}})$
\STATE Initialize $\mathcal{T}_i \gets \varnothing$

\FORALL{$T_i \in \mathcal{B}$}
    \STATE $\mathcal{C}_i
    \gets
    \mathcal{A}_{\text{ctrl}}(\mathcal{I}_i,q_i,\mathcal{T}_{<i},\mathcal{S})$
    \STATE $\mathcal{T}_i \gets \texttt{InnerLoopEdit}(\mathcal{C}_i)$
\ENDFOR

\STATE $\mathcal{T}_{\text{comp}} \gets \bigcup_{i=1}^{M}\mathcal{T}_i$

\STATE $\mathcal{G}_{\text{conf}}
\gets
\mathcal{A}_{\text{dia}}(\{\mathcal{T}_i\}_{i=1}^{M})$

\IF{$\mathcal{E}_{conf} \neq \varnothing$}

    \STATE $\mathcal{R}
    \gets
    \texttt{RegionalDecompose}(\mathcal{G}_{conf},\mathcal{G}_{task})$

    \REPEAT

        \FORALL{$\kappa_r \in \mathcal{R}$}
            \STATE $q_{\kappa_r}^{rep}
            \gets
            \mathcal{A}_{neo}(\kappa_r)$
            \STATE Re-edit segments in $\kappa_r$ using InnerLoop
        \ENDFOR

        \STATE $\mathcal{G}_{conf}
        \gets
        \mathcal{A}_{dia}(\{\mathcal{T}_i\})$

        \STATE Update $\mathcal{R}$ by merging conflicted regions

    \UNTIL{$\mathcal{E}_{conf}=\varnothing$ or max iteration}

\ENDIF

\STATE $\mathcal{T}^{*} \gets \texttt{GlobalRefine}(\{\mathcal{T}_i\})$

\STATE \textbf{return} $\mathcal{T}^{*}$

\end{algorithmic}
\end{algorithm}

\begin{algorithm}[t]
\caption{\methodname{} Inner-Loop Segment-Level Editing}
\label{alg:inner_loop}
\begin{algorithmic}[1]
\REQUIRE Context bundle $\mathcal{C}_i$
\ENSURE Segment timeline $\mathcal{T}_i$

\STATE Extract $(q_i,\mathcal{I}_i,\mathcal{S})$ from $\mathcal{C}_i$

\STATE $\{q^{ret}_{i,t}\}_{t=1}^{n_r}
\gets
\mathcal{A}_{ret}(q_i)$

\FORALL{$q^{ret}_{i,t}$}
    \STATE $\mathcal{U}_{i,t}
    \gets
    \texttt{RetrieveClips}(q^{ret}_{i,t},\mathcal{S})$
\ENDFOR

\STATE $\{\hat{u}_{i,k}\}
\gets
\texttt{RankAndFilter}(\cup_t \mathcal{U}_{i,t})$

\STATE $\mathcal{T}_i^{(0)}
\gets
\mathcal{A}_{rc}(\{\hat{u}_{i,k}\},q_i,\mathcal{I}_i)$

\FOR{$t=0$ to MaxIter}

    \STATE $(pass,d_i^{(t)})
    \gets
    \texttt{Diagnose}(\mathcal{T}_i^{(t)})$

    \IF{$pass$}
        \STATE \textbf{return} $\mathcal{T}_i^{(t)}$
    \ENDIF

    \STATE $\mathcal{T}_i^{(t+1)}
    \gets
    \mathcal{A}_{ar}(\mathcal{T}_i^{(t)},d_i^{(t)})$

    \IF{NeedMoreClips$(d_i^{(t)})$}
        \STATE Retrieve additional clips
        \STATE Reconstruct rough cut
    \ENDIF

\ENDFOR

\STATE $\mathcal{T}_i
\gets
\arg\max_{\mathcal{T}}
s_i(\mathcal{T})$

\STATE \textbf{return} $\mathcal{T}_i$

\end{algorithmic}
\end{algorithm}

\section{MVEBench}
\label{app:MVEBench}

\subsection{Details of Taxonomy}
\label{app:details-of-taxonomy}
We introduce the details of the benchmark taxonomy here.

\paragraph{Types of Mashup Videos.}

The collected mashup videos can be broadly categorized into two editing styles: on-beat mashups and story-driven mashups. On-beat mashup videos primarily focus on rhythmic synchronization between video shots and music beats. These videos typically lack explicit narrative structures and instead arrange visually impactful shots to match the rhythm and energy progression of the music. Examples include action-movie highlight mashups in which shots are tightly aligned with musical beats. In contrast, story-driven mashup videos emphasize narrative coherence and storytelling. Editors reorganize or reinterpret source footage to construct a new storyline or summarize the original plot. Typical examples include misleading montage videos or condensed storytelling edits that rearrange scenes to produce alternative interpretations. These two types of mashups emphasize different evaluation criteria: on-beat mashups prioritize rhythmic alignment and visual dynamics, whereas story-driven mashups focus more on narrative coherence and semantic consistency.

\paragraph{Difficulty Levels Based on Music Length.}

Music length significantly affects the complexity of the editing task. Longer music tracks require longer-horizon planning, more segment-level decisions, and stronger global coordination across the timeline. They also introduce more potential cross-segment conflicts during timeline construction. To systematically evaluate different levels of task complexity, we categorize MVEBench instances into three difficulty levels based on music duration. Easy tasks correspond to music tracks shorter than approximately 30 seconds, where the music typically contains only one to three structural segments and editing decisions mainly involve local alignment. Medium tasks correspond to music durations between approximately 30 and 90 seconds. These tracks usually contain several structural segments and require moderate global planning to resolve interactions between neighboring segments. Hard tasks correspond to music tracks longer than 90 seconds, which often contain more than ten structural segments and require long-horizon planning as well as conflict-aware coordination across the entire timeline. The duration thresholds are not strictly fixed. Instead, the final difficulty label is determined by human annotators who consider both music structure and editing complexity. In practice, most collected mashup videos have durations of approximately three to four minutes. To construct balanced difficulty levels, annotators manually segment long mashup videos according to music structure and editing boundaries. For example, a four-minute mashup video may be divided into several shorter editing tasks with different difficulty levels. All segmentation operations are manually verified by annotators to ensure consistency.

\paragraph{User Intent Design.}

Another key component of MVEBench is the design of user input. Since music-guided mashup video editing is an open-ended task, a fixed ground-truth output is impractical. Instead, we design two levels of user intent descriptions to evaluate different capabilities of editing agents. The first type is \emph{general-level user intent}, which provides only a high-level description of the desired video. For example, the user may request to create a high-energy and positive mashup video for \emph{Harry Potter} based on the given music. Such inputs specify the overall goal but do not impose detailed structural constraints, requiring the editing agent to autonomously perform global planning, shot retrieval, and timeline construction. The second type is \emph{medium-level user intent}, which provides a coarse script aligned with the music timeline in addition to the overall goal. For example, the user description may specify the desired scene types or emotional transitions for different music segments. This setting evaluates the ability of an editing framework to follow structured instructions while resolving cross-segment conflicts. We intentionally avoid overly detailed user instructions that specify every individual shot. Such a setup would reduce the task to a simple video retrieval problem and would not capture the core challenges of mashup video editing, including planning, composition, and maintaining global timeline coherence.

\section{Agentic Evaluation Framework}
\label{app:agent-evaluation-framework}
\subsection{Motivation}
Evaluating music-guided mashup video creation is inherently challenging due to the open-ended nature of the task. 
Given the same music track, prompt, and source video pool, multiple edited results may all be reasonable yet stylistically different. 
For example, for a prompt such as \emph{``create a high-energy Harry Potter mashup that conveys a joyful magical life''}, one editor may focus on battle scenes and spell casting, while another may emphasize vivid daily-life moments at Hogwarts. 
Both interpretations can satisfy the prompt, making simple reference-based or single-metric evaluation inadequate.
A natural alternative is to use a strong multimodal large language model (MLLM) as a judge. 
However, directly asking a monolithic model to score a long mashup video is often unreliable. 
Such videos contain dense temporal structures, fine-grained transitions, and cross-segment dependencies that are difficult to evaluate in a single pass. 
Moreover, several evaluation dimensions in this task require explicit grounding in external signals, such as music beats, emotion trajectories, shot boundaries, and shot-level semantics.

\subsection{Reasons for excluding objective and low-level metrics}
\label{app:reasons-exclude-objective-metrics}
We acknowledge that relying solely on an agent-as-a-judge strategy has limitations due to the potential unreliability of the backbone LLM. However, in our task it is difficult to directly apply objective or low-level metrics such as beat-hit rates or embedding similarities. Take beat-hit as an example: although \emph{rhythm alignment} is an important evaluation dimension for on-beat video editing, it cannot be reliably measured by simply counting how often shot boundaries coincide with detected musical beats. Effective rhythm alignment also depends on which beats are selected, how edits match musical phrase structures, and whether visual transitions correspond to the audio dynamics. As a result, a video that cuts mechanically on every detected beat may obtain a high beat-hit score while still appearing unnatural or poorly paced.

Similar issues arise for other evaluation dimensions such as emotion alignment, story completeness, character continuity, and instruction following. These criteria involve compositional and contextual judgments across multiple segments rather than frame-level similarity. Low-level metrics such as embedding similarity or synchronization statistics can capture only partial signals and often fail to reflect whether the edited video truly satisfies the intended criterion.

Therefore, while the evaluation dimensions themselves are meaningful, their quality is difficult to quantify using simple objective formulas. We thus report these criteria through judge-based evaluation instead of handcrafted automatic metrics. Although imperfect, this approach better captures the high-level and holistic properties required for evaluating music-video mashup editing.

\subsection{Details of Evaluation Criteria}
\label{app:details-evaluation-criteria}

\paragraph{Agentic evaluation for on-beat mashup videos.}
For on-beat mashup evaluation, our framework first invokes a music analysis agent to extract beat timestamps, downbeat positions, tempo-related patterns, music energy curves, and segment-level emotional attributes.
In parallel, a video analysis agent detects shot boundaries and transition timestamps from the generated mashup video, and reconstructs the temporal editing structure.
These outputs are aligned into a structured timeline representation that explicitly records the correspondence between music beats and visual transitions.
Based on this representation, a reasoning agent evaluates \textbf{Rhythm Alignment} by analyzing whether shot changes, salient motion events, or segment transitions occur at musically appropriate positions.

To evaluate \textbf{Emotion Alignment}, the framework further performs segment-level semantic understanding.
Specifically, a multimodal captioning agent generates descriptions for individual shots or short temporal spans, and a summarization agent aggregates them into segment-level semantic summaries.
These summaries are then combined with music-side evidence, including segment emotion labels and energy profiles, to form an evidence package describing both the intended emotional trajectory of the music and the actual visual content presented by the mashup.
The judge model reasons over this grounded evidence to determine whether the visual composition matches the affective tone, intensity, and progression of the music.

Finally, \textbf{Overall Quality} is evaluated using all collected evidence, including temporal alignment signals, segment summaries, global video structure, and prompt relevance.
Rather than treating overall quality as a vague subjective score, our framework asks the judge model to assess whether the final mashup is globally coherent, visually natural, musically compatible, and reasonably faithful to the prompt.
Because this score is conditioned on structured evidence collected by upstream agents, it better reflects the true compositional quality of the mashup than direct end-to-end judgment from raw video alone.

\paragraph{Agentic evaluation for story-driven mashup videos.}
For story-driven mashup evaluation, the framework emphasizes higher-level semantic and temporal coherence.
As in the on-beat setting, the framework first performs music analysis and video structure analysis.
However, the main goal here is not beat-level synchronization, but narrative organization and character consistency across the entire mashup.

To evaluate \textbf{Story Completeness}, a semantic analysis agent first captions shots and summarizes temporally adjacent shots into segment-level events.
These event summaries are then organized into a coarse story timeline.
Conditioned on the prompt and the reconstructed timeline, the reasoning agent evaluates whether the video presents a complete and understandable story arc, including whether the beginning, development, and conclusion are sufficiently represented, and whether the sequence of events is logically connected.
This criterion is particularly important for long-horizon editing, where local shot quality alone does not guarantee a globally meaningful narrative.

For \textbf{Character Continuity}, the framework uses a character-aware analysis agent to track major entities across segments.
This agent leverages visual descriptions, identity-sensitive multimodal cues, and cross-segment semantic consistency signals to determine whether the main characters remain stable throughout the mashup.
The judge model then reasons over these observations to assess whether the edited video preserves character identity, avoids abrupt or confusing substitutions, and maintains consistent character-centric progression over time.

For \textbf{Overall Quality} in story-driven videos, the framework again aggregates all evidence produced by the preceding agents, including event summaries, story timeline structure, character continuity cues, prompt relevance, and global compositional information.
The final judge considers whether the mashup is not only locally understandable, but also globally well-paced, semantically coherent, and aesthetically reasonable as a complete story-driven video.

\begin{table*}[t]
\centering
\small
\caption{Ablation of global--local coordination modules on \textbf{story-driven} mashup video creation. Efficiency is reported as normalized processing efficiency, with higher being better.}
\label{tab:ablation_components_story}
\begin{tabular}{lcccccccc}
\toprule
Variant
& Preventive
& Region Decomp.
& Bottom-up Negot.
& \makecell{Story\\Completeness$\uparrow$}
& \makecell{Character\\Continuity$\uparrow$}
& \makecell{Instruction\\Following$\uparrow$}
& \makecell{Overall\\Quality$\uparrow$}
& \makecell{Efficiency$\uparrow$} \\
\midrule
Only Preventive Controller & \checkmark &  &  & 3.20 & 3.10 & 3.38 & 3.07 & \textbf{1.18} \\
w/o Bottom-up Negotiation  & \checkmark & \checkmark &  & 3.45 & 3.37 & 3.63 & 3.28 & 1.10 \\
w/o Region Decomposition   & \checkmark &  & \checkmark & 3.56 & 3.48 & 3.72 & 3.34 & 0.76 \\
w/o Preventive Controller  &  & \checkmark & \checkmark & 3.69 & 3.59 & 3.82 & 3.38 & 0.95 \\
Full \methodname{}                & \checkmark & \checkmark & \checkmark & \textbf{3.75} & \textbf{3.68} & \textbf{3.95} & \textbf{3.42} & 1.00 \\
\bottomrule
\end{tabular}
\end{table*}

\begin{figure*}[t]
    \centering
    \includegraphics[width=0.95\linewidth]{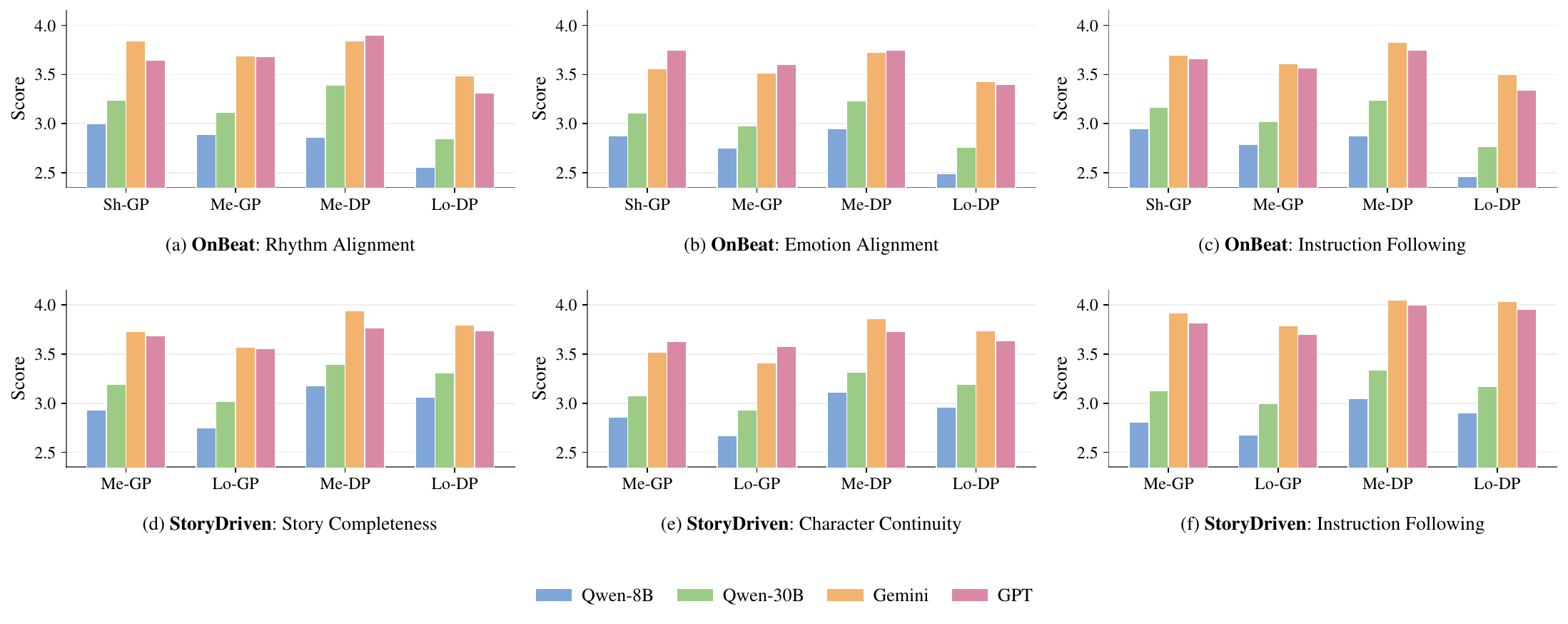}
    \caption{Evaluation scores across different task configurations..}
    \label{fig:difficulty-results}
\end{figure*}

\subsection{Benefits of Agent-as-a-judge}
\label{app:genefit-aaaj}
Compared with directly prompting a single MLLM to score long videos, the proposed framework explicitly decomposes evaluation into \emph{perception} and \emph{reasoning} stages. 
This structured design grounds the final judgment in interpretable intermediate evidence and improves robustness when evaluating long-form mashup videos with complex temporal structures.

\section{Human Evaluation and Judge Consistency}
\label{app:judge-consistency}

\subsection{Ablation Studies}

Table \ref{tab:ablation_components_story} shows the ablation study on Story-driven tasks. Similar to the On-beat results shown in Table \ref{tab:ablation_components_onbeat}, \methodname{} with complete components outperforms all other baselines without one or more modules.

\end{document}